\documentclass[12pt] {article}
\usepackage{psfig}
\usepackage{color}
\begin{document}
\hrule
\begin{center}
{\Huge \bf
Relationship between horizontal Flow Velocity  \\  \&  Cell lifetime for
supergranulation from SOHO Dopplergrams}
\vspace{0.5cm}

\hrule
 {\Large U.Paniveni$^1$, V.Krishan$^1$, Jagdev Singh$^1$, R.Srikanth$^2$ \\   
 $^1$ Indian Institute of Astrophysics, Koramangala, Bangalore-560034, India\\
$^{2}$ Optics group, Raman Research Institute\\
 Bangalore-560080, Karnataka, India.}
\end{center}

{\bf  ABSTRACT}
~\\
\hrule
~\\
{\bf
        A study of $50$ supergranular cells obtained from SOHO Dopplergrams was
undertaken in order to investigate the relationship between the lifetime ($T$)
and the horizontal flow velocity  ($v_h$)  of the cells. For this sample we find
that the two parameters are correlated with a relation $v_h\propto T^{0.5}$ and 
$T$ is identified with the eddy turn-over time. This is in agreement with 
the turbulent convective model of the solar atmosphere where the velocity 
spectrum of supergranular field given by '$v_h \propto L^{1/3}$' can be  
identified with the Kolmogorov spectrum for the eddy size $L$.}
~\\

{\bf
INTRODUCTION}
~\\
\hrule
~\\
 Convection is the chief mode of transport of heat in the outer envelopes of cool stars such as 
the sun. The convection zone which lies in the sub-photospheric layers of the sun has a 
thickness of about 30\% of the solar radius. Here the opacity is so large that energy is carried by turbulent motions rather than by photon diffusion. The convective motions on the sun are 
characterized by two prominent scales: the granulation with a typical size of 1000 km and the 
supergranulation with a typical size of 30000 km. The supergranules are regions of horizontal 
outflow along the surface, diverging from the cell centre and subsiding flow at the cell 
borders. Such outflowing regions always show velocity of approach to the observer on the side 
close to the centre of the disk and velocity of recession on the side towards the limb. Near the centre of the disk where the horizontal outflows are transverse to the 
line-of-sight, there is less dopplershift and so the image is almost uniformly grey.
These high photospheric large convective eddies sweep up any shreds of 
photospheric magnetic fields in their path from the declining active regions 
into the boundaries of the cell, where they produce excess heating,
resulting in the chromospheric network. The approximate lifespan of a 
supergranular cell is $24$ hours. Broadly speaking supergranules are 
characterized by the three parameters namely length $L$, lifetime $T$ and 
horizontal flow velocity $v_h$. The interrelationships amongst these parameters 
can shed light on the underlying convective processes. Cells of a given size 
associated with remnant magnetic field regions live longer than those in the 
field-free regions (Singh et al. 1994).The lifetime of network 
cells was found to be larger for active-region cells as compared to that of 
quiet-region cells (Raju, Srikanth and Singh 1998). Diffusion-like dispersion of
the magnetic flux is the dominant factor in the large scale evolution of the 
network. The lifetime of a supergranular cell is found to depend 
on the size of the cell and is larger for bigger cells (Srikanth et al. 1999). 
Convective motion and magnetic inhibition of motion are both stronger 
in active regions, thereby leading to similar speeds in all regimes (Srikanth, 
Singh and Raju, 1999). A positive correlation between horizontal flow velocity 
and cell size of a supergranular cell has been established recently by Krishan 
et al. (2002). The corresponding dependence of lifetime of the supergranular 
cell on its horizontal flow velocity is expected to be $v_h\propto T^{0.5}$ 
where the  eddy turn-over time i.e. the crossing time for plasma from centre to 
edge can be estimated from the relation $T=L/v_h$ with $L$ as the distance from 
the centre to the edge of the cell and $v_h$ is the peak horizontal flow 
velocity of the cell. In this paper we report on this possible interrelationship between horizontal flow velocity and cell lifetime for supergranules.

~\\

{\bf Source of data}
~\\
\hrule
~\\ 
We analysed $33$ hour data of full disc Dopplergrams obtained on 28th and 29th 
June 1996 by the Michelson Doppler Interferometer (MDI) on board the solar and 
Heliospheric observatory (SOHO) (Scherrer et al. 1995).

~\\
\begin{figure}[hd]
\centerline{\psfig{figure=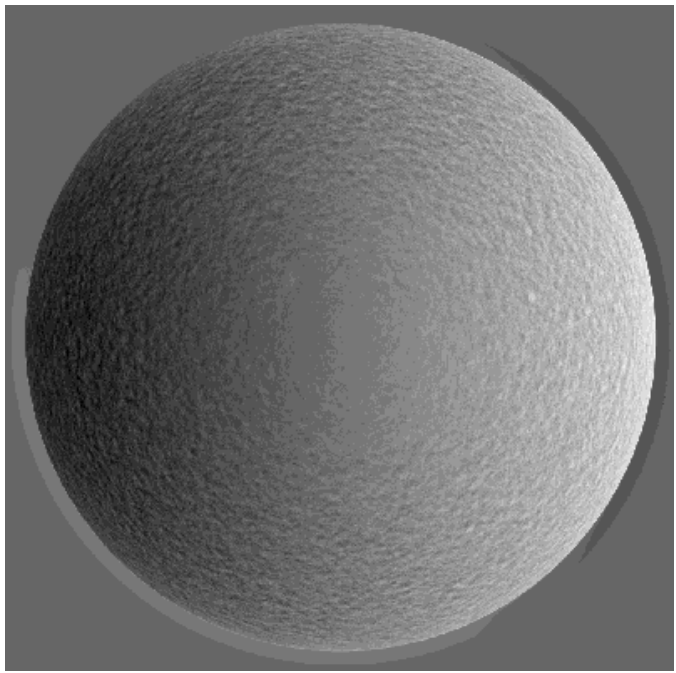,width=15cm}}
\caption{Processed SOHO Dopplergrams}
\end{figure}

\newpage

{\bf Data Analysis}
~\\
\hrule
~\\
The SOHO full disc Dopplergram data has been obtained with a resolution of 
2$^{\prime\prime}$ which is twice the granular scale. Further, the Dopplergrams 
are time averaged over intervals of 10 min, which is about twice the 5-minute 
period of oscillations. Thus the signal due to granular velocity is averaged 
out. Similarly the contributions due to p-mode vibrations are  
reduced after the time averaging. Our analysis rests on the implicit belief that
time averaging removes noise significantly. Noise is reduced in our data 
considerably with 10-minute integration time, as judged from visual inspection 
and also as seen in typical  supergranular velocity profile for our data (Figure 2). After the 
averaging, the supergranular network is brought out with a fair 
clarity. This procedure yielded usually six images per hour of the data. 
Corrections due to solar rotation are applied to the Dopplershifts. Fifty well 
accentuated cells lying between 15$^{\circ}$ and 60$^{\circ}$ angular distance 
from the disc centre were selected. Restricting to the above mentioned angular 
distance limits helps us discount weak supergranular flows as well as 
foreshortening effects.

~\\

{\bf Data Processing}
\hrule
~\\
{\bf Supergranular cell speed}
\hrule
~\\

 The line-of-sight velocities in the dark/bright region of the cells are 
directly read off from the velocity scan. Among them the first three highest 
velocity read-outs and the last three least velocity read-outs were selected. 
The maximum cell velocity is then determined as the average of the former three 
values minus the average of the latter three. This furnishes a simple way to 
assign a peak horizontal flow velocity $v_h$ to a given cell that is independent of large scale 
velocity gradients. To see this let us write $v_{\rm max} = |v_h| + v_{\oplus}$ where 
$v_{\oplus}$ represents contributions due to large scale 
gradients in the velocity field and $v_{\rm min} = -|v_h| + v_{\oplus}$. Then 
half the difference of $v_{\rm max}$ and $v_{\rm min}$ is the required peak 
horizontal velocity $v_h$. Three pairs of values were chosen to add robustness.
Now, it is a fact that even a randomly noisy velocity field can, by our method 
of choice of three largest velocities, be biased to yield some spurious 
relations (eg.,larger cells showing larger velocity). Hence it is of 
considerable importance to be certain that the data is not noisy at the level of interest. The 
ten-minute time-averaging removes noise to a level sufficient for 
our  purpose. This was clear from a visual inspection of the images. More 
specifically, the selected three peak positive velocity points and the three 
peak negative points are not spiky and  fit in smoothly with the surrounding 
flow pattern and hence the chosen peaks are with very high probability part of 
the velocity profile and not noise. This is depicted for a typical cell profile in Figure 2.
Dopplergrams give us the line-of-sight velocity component. Geometrically it 
has a contribution from the local horizontal flow field $v_h$ and vertical flow 
field $v_v$. Normally, the vertical component can be ignored because the 
convective upwelling, concentrated towards the cell centre, and the downflows,
along the periphery of the cell, are typically much smaller. However, 
regions with a considerable vertical component of velocity are not improbable.
For example, the upflow regions can be as broad as 10$^{\prime\prime}$ 
(K\"uveler 1983). More importantly, it follows from basic trignometric arguments
that our method of velocity selection will tend to pick up
the three largest positive values from approaching flows with a considerable
upflow component, and the (magnitudinally) largest negative values from receding
flows with a considerable downward component.
Therefore, in order to account for this residual $v_v$ contribution (where we
treat upflows and downflows symmetrically for simplicity), we 
need to make an assumption about the relation connecting $v_h$ and $v_v$. 
The velocity derived from our analysis is based on this assumption. 
In the reported literature, $v_h$ is known to be larger than $v_v$. This is also supported by 
mass conservation law (Krishan 1999 and Krishan et al. 2002). 
Direct inspection of the disk center yields vertical velocities of about $200$ m s$^{-1}$. For 
$v_h = 539.15$ m s$^{-1}$, which we obtain as mean value, this 
implies a value closer to $r \equiv v_v/v_h = 0.4$. Hence this value of $r$ is 
adopted for the present analysis.
~\\

{\bf Supergranular cell lifetime}
~\\
\hrule
~\\
 The $33$ hour data is spread over $198$ frames with a $10$ minute interval 
between the consecutive frames. Only those cells which appear and disappear 
within the chosen period of $33$ hours are considered. This excludes cells 
already present in the first frame and those still present in the last frame. 
Thus the selected cells were born a few frames after the first. A particular 
supergranular cell thus identified is tracked down the successive frames until 
it disappears completely in a particular frame. Lifetime is identified to be the time interval between its first appearance and  final disappearance.

~\\
~\\
{\bf Results}
~\\
\hrule
~\\

 A large dispersion in the supergranular lifetime and horizontal velocity is 
noted. 
 The main results pertaining to the  maximum, the minimum, the mean and the 
standard deviation for horizontal flow velocity $v_h$ and cell lifetime $T$ are 
summarized in Table 1.(Computed using $r = 0.4$.)

\begin{table}[h]
\begin{center}
\begin{tabular}{lllll}
\hline
   &  Max &   Min  &     Mean &     $\Sigma$  \\
\hline
$T$ (hr) &   32.00  &    16.00  &         22.00  &       3.0 \\
\hline
$v_h$(m/s) & 661.06    &  402.33     &     539.15   &      1.8 \\
\hline
$L$(Mm)    &42.79      &  17.63      &     27.42     &     5.75 \\
\hline
\end{tabular}
\caption{Maximum, minimum, mean and standard deviation  for cell lifetime 
 ($T$) and cell peak horizontal flow velocity ($v_h$) and cell size($L$).}
\end{center}
\end{table}

 In  previous studies cell lifetimes were derived by cross-correlation 
technique (Raju, Srikanth and Singh 1998a). The analysis was based on diffusion 
co-efficients of the network magnetic elements and was
identified as a 'diffusion lifetime'. Estimation of lifetime by visual 
inspection method rests on the crossing time for plasma from centre to edge of 
the cell (Krishan 1999). Hence visual inspection is expected to
lead to the eddy turn-over time given by $T = L/v_h $.
Visual inspection is a fairly foolproof method, though laborious. The sample is 
quite small but is characteristic for different epochs and regions.

Figure (3) presents a  plot of square of the  peak horizontal flow velocity with $r=0.4$ and 
cell lifetime.
 A powerlaw of the form
\begin{equation}
\label{curvefit}
v_h= CT^{\alpha}
\end{equation}
was fitted to the data using the least squares method. For $r=0.4$,
we find $C = 0.001$  and $\alpha$ =$ 1.09$.Its intercept gives the value of 
$\epsilon\approx 1.37\times 10^{-6}km^{2}s^{-3}$. Figure (3) clearly shows that 
the two parameters are well correlated and the correlation co-efficient between 
$v_h^{2}$ and $T$ is about 0.78.

~\\

{\bf Discussion and conclusions}
~\\
\hrule
~\\
 In an earlier study (Krishan 1991), it was suggested that the granulation and supergranulation 
are the result of energy cascading processes in a turbulent
medium. Recently we showed that the velocity spectrum of the supergranular field
very closely agrees with Kolmogorov's spectrum $v_h =\epsilon^{1/3}L^{1/3}$ 
(Krishan et al.2002) where $\epsilon$ is the energy 
injection rate. Defining the eddy turn-over time to be the lifetime of the 
cell, we can write $T = L/v_h$. Combining with the Kolmogorov spectrum $v_h = 
\epsilon^{1/3}L^{1/3}$, we find 
\begin{equation}
 v_h = \epsilon^{1/2} T^{1/2}
\end{equation}

Comparing (1) and (2) we see that $C=\epsilon^{1/2}$ and hence $\epsilon = 
10^{-6} km^{2}s^{-3}$ which is very close to the value of $\epsilon$ obtained 
earlier by us (Krishan et al. 2002) from the velocity $v_h$ and size $L$ 
relationship 
with different data. Using the data of peak 
horizontal flow velocity of 50 supergranular cells and their sizes, 
we have plotted the data of  $v_h$ 
against  $L$ as shown in Figure (4). Its intercept gives the value of 
$\epsilon \approx 5.26\times 10^{-6}$ km$^{2}$s$^{-3}$ and a slope of 0.33. The 
correlation co-efficient for $v_h$ and $L^{1/3}$ of 0.68 is obtained. Thus we 
conclude that the supergranular velocity field is well accounted for by the 
Kolmogorov spectrum $v_h = \epsilon^{1/3}L^{1/3}$ with eddy turn-over time 
realised from $v_h = \epsilon^{1/2}T^{1/2}$ . A comparison of the theoretical 
energy spectrum of granulation ($\propto K^{-5/3}$), mesogranulation ($\propto 
K^{-1}$) and supergranulation ($\propto K^{-5/3}$) (Krishan 1991 ; 1996) with 
the observed energy spectrum of granulation ($\propto K^{-5/3}$), 
mesogranulation ($\propto K^{-0.7}$) and supergranulation ($\propto K^{-5/3}$) 
(Malherbe et al.1987; Zahn 1987; Keil et al.1994) appears to  point at the 
phenomenon of the inverse cascade of energy operating in the solar convective 
turbulence.

~\\
{\Large \bf Acknowledgement}
~\\
\hrule
~\\
{\bf We thank Dr. P.H. Scherrer and the SOHO consortium for providing us 
with MDI/SOI data}.

\onecolumn
\small
\begin{figure}
\vspace{102pt}
\centerline{\psfig{figure=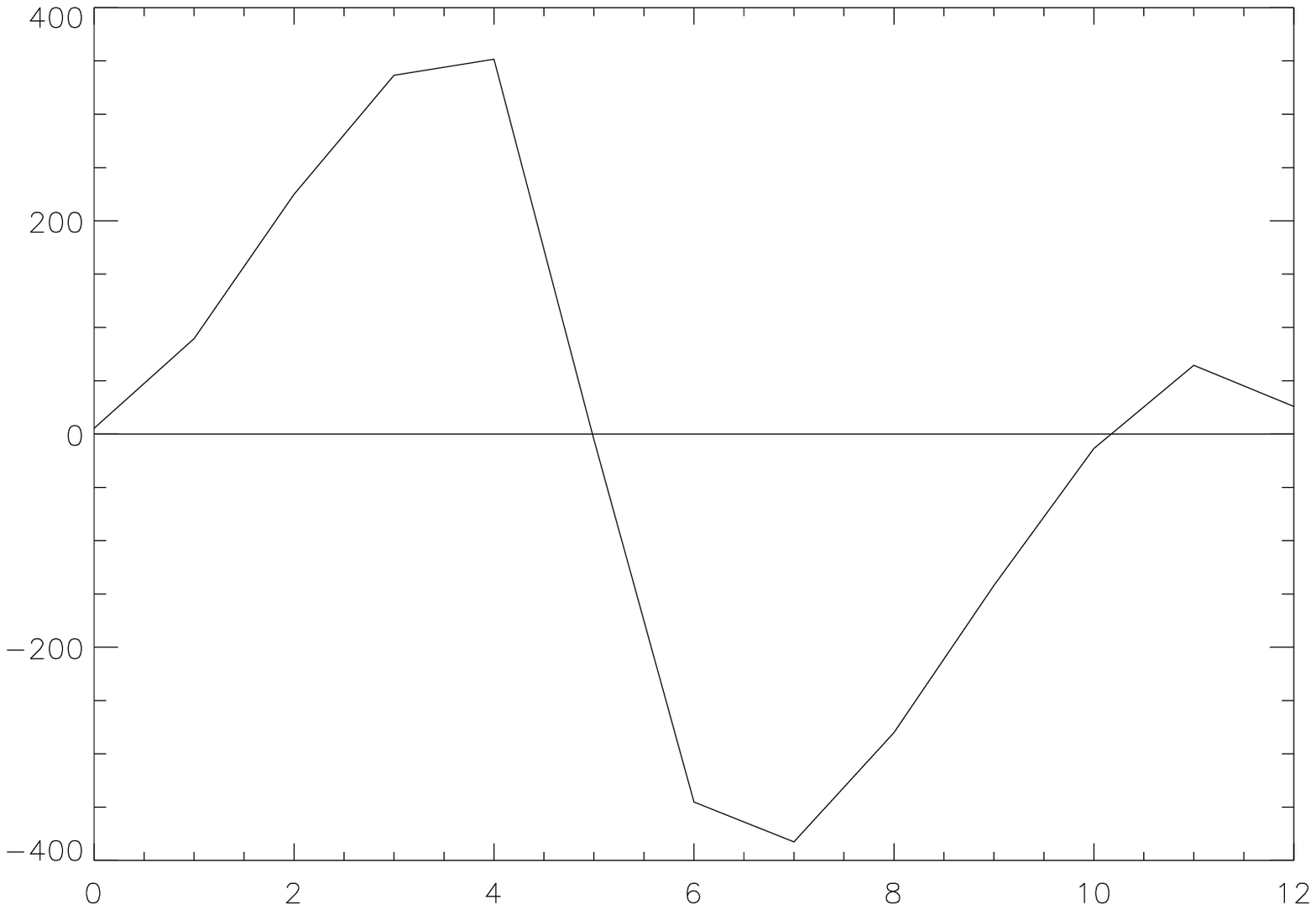}}
\caption{Profile of the line of sight velocity component $v_L$ in 
$m s^{-1}$ on the y-axis against cell extent $x$
in pixels on the x-axis}. \end{figure}

\begin{figure}
{\centerline{{\psfig{figure=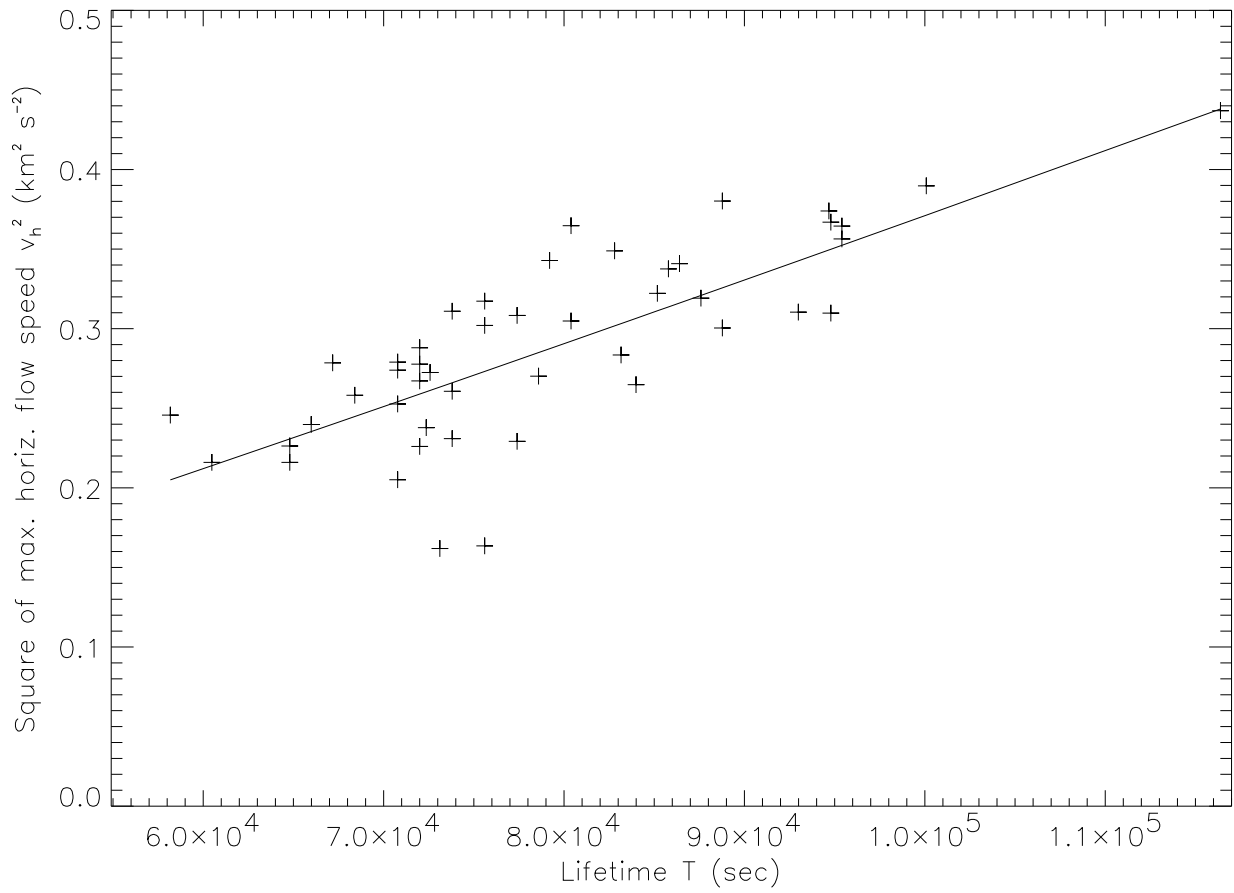}}}
\caption{Plot of square of Peak horizontal velocity of the supergranular 
cell $v_h^{2}$ against cell lifetime.The measured values are 
represented by the plusses. The line is based on a least squares fit 
to Eq. (\ref{curvefit}).} }
\end{figure}

\begin{figure}
{\centerline{\psfig{figure=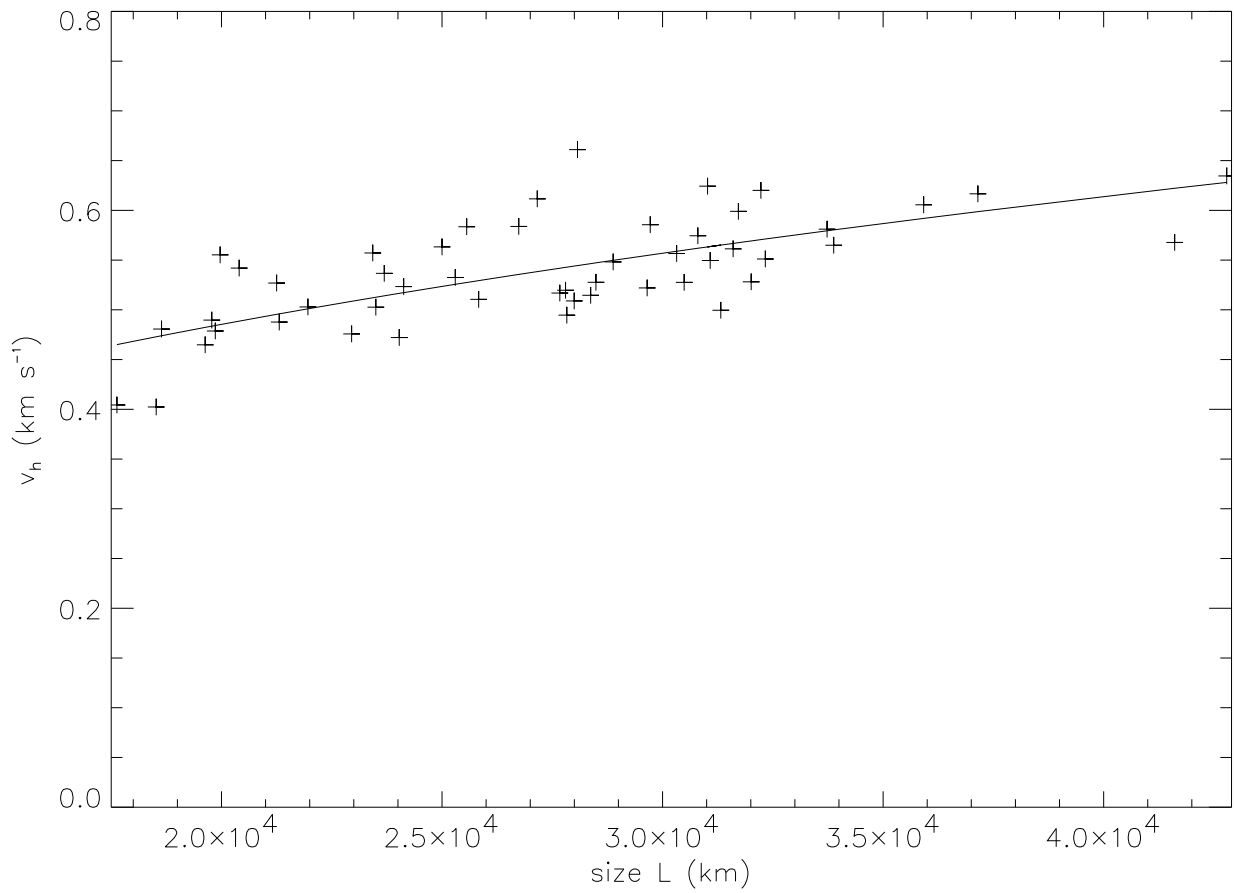}}
\caption{Plot of peak horizontal velocity $v_h$ against cell size $L$. The 
measured values are represented by plusses. The line is based on least 
squares fit to the equation $v_h = \epsilon^{1/3} L^{1/3}$}}

\end{figure}

\end{document}